\documentclass[]{rsos}

\usepackage{color}
\usepackage[table,dvipsnames]{xcolor}
\definecolor{mygray}{gray}{0.6}

\newcommand{\arxiv}{arXiv}

\begin{document}

\title{The impact of the COVID-19 pandemic on academic productivity}

\author{
A. R. Casey$^{1,2}$,  I. Mandel$^{1,3,4}$ and P. K. Ray$^{5}$}

\address{$^1$School of Physics \& Astronomy, Monash University, Clayton 3800, Victoria, Australia\\
$^2$ARC Centre of Excellence for Astrophysics in Three Dimensions (ASTRO-3D), Australia\\
$^3$OzGrav, Australian Research Council Centre of Excellence for Gravitational Wave Discovery, Australia\\
$^4$Institute of Gravitational Wave Astronomy and School of Physics and Astronomy, University of Birmingham, Birmingham, B15 2TT, United Kingdom\\
$^5$Department of Mathematics, Imperial College London, London, United Kingdom
}

\subject{xxxxx, xxxxx, xxxx}


\corres{Andrew R. Casey\\
\email{andrew.casey@monash.edu}}

\begin{abstract}
`Publish or perish’ is an expression describing the pressure on academics to consistently publish research to ensure a successful career in academia. 
With a global pandemic that has changed the world, how has it changed academic productivity? 
Here we show that academics are posting just as many publications on the arXiv pre-print server as if there were no pandemic: 168,630 were posted in 2020, a +12.6\% change from 2019 and $+1.4\sigma$ deviation above the predicted 162,577 $\pm$ 4,393. 
However, some immediate impacts are visible in individual research fields. 
Conference cancellations have led to sharp drops in pre-prints, but laboratory closures have had mixed effects. 
Only some experimental fields show mild declines in outputs, with most being consistent on previous years or even increasing above model expectations. 
The most significant change is a 50\%~increase ($+8\sigma$)  in quantitative biology research, all related to the COVID-19 pandemic.
Some of these publications are by biologists using arXiv for the first time, and some are written by researchers from other fields (e.g., physicists, mathematicians). 
While quantitative biology pre-prints have returned to pre-pandemic levels, 20\% of the research in this field is now focussed on the COVID-19 pandemic, demonstrating a strong shift in research focus.
\end{abstract}

\maketitle


\noindent Peer-reviewed publications are the most common measure of productivity in academia. The \arxiv\cite{Ginsparg:2011} ({https://arxiv.org}) is a distribution service for research publications before they are printed in a journal (i.e., a pre-print). A pre-print on \arxiv\ does not ensure that the contents have passed peer-review, but most material on \arxiv\ eventually goes through peer-review because it is now standard in many research fields to post to \arxiv\ either during or after the peer-review process \cite{Lariviere:2014}. For this reason, the number of pre-prints posted to \arxiv\ approximates the number of peer-reviewed publications written in these fields.

Here we make quantitative comparisons on the number of pre-prints posted to \arxiv\ before and during the pandemic. While we investigate the impact that the COVID-19 pandemic has had on different fields of research, we highlight that we are unable to identify or differentiate between authors and communities whose productivity has been significantly harmed by the pandemic, and those who were largely unscathed. The COVID-19 pandemic has impacted the population in unequal ways\cite{Nicola:2020,Chu:2020,Viglione:2020,Gewen:2020,IbnMohammed:2021,King:2021}. Generational inequality, career stage, personal circumstances, carer responsibilities, work environments, places of employment, and many other factors, all significantly contribute to the disproportionate and unequal impact the pandemic may have had on a scientist's capacity to conduct research. While this limited analysis does not address these issues, it is important to consider that if a community has not yet demonstrated a significant change in productivity, many researchers have faced significant challenges and suffered physically, mentally, emotionally, and professionally.

\begin{figure}[!h]
\includegraphics[width=0.95\linewidth]{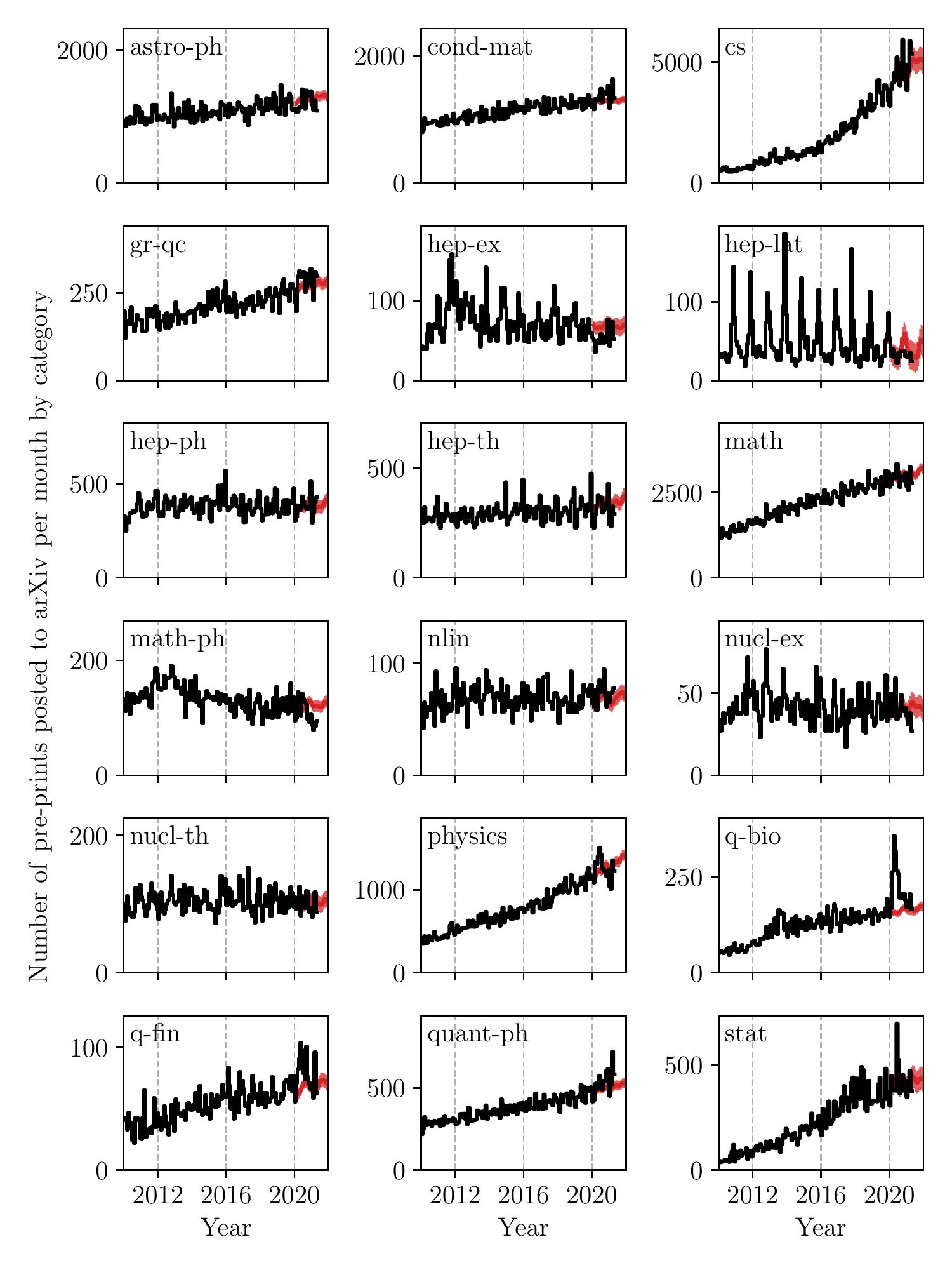} 
\caption{There has been no significant change in monthly \arxiv\ pre-prints (black) since the COVID-19 pandemic began. The expected number of pre-prints from our Gaussian Process model is shown in red, and the red region shows the uncertainty in that expectation. The dashed light grey line shows predictions from an autoregressive integrated moving average model.}
\label{fig:1}
\end{figure}

We retrieved metadata for 1,475,914 pre-prints posted to \arxiv\ between 1 April 2007 and 31 May 2021. The metadata includes the creation date, research field(s), title, author name(s), abstract, and other miscellaneous information\cite{Clement:2019}. There is an increasing number of pre-prints posted to \arxiv\ each year in nearly every field (Figure~\ref{fig:1}). These long-term trends are relatively predictable from year to year, allowing us to quantify any change in academic productivity due to the COVID-19 pandemic. We used the number of publications from January 2015 to December 2019 to predict the expected number of monthly pre-prints in each field after January 2020. We modelled the number of monthly pre-prints in each field with a straight line to capture long-term (linear) trends, and used a Gaussian Process with a quasi-periodic kernel function\cite{Rasmussen:2006,Ambikasaran:2014} to capture seasonal variations (Figure~\ref{fig:1}). In nearly all research fields the number of pre-prints posted since January 2020 agree excellently with the model predictions (Table~1), indicating no collective impact on academic productivity due to the COVID-19 pandemic. The total number of \arxiv\ pre-prints in 2020 across all fields (168,630) exceeds our model predictions ($162{,}577 \pm 4{,}393$; $+1.4\sigma$).

Border closures and travel restrictions have forced many academic conferences to be held online-only, or to be cancelled. Hosting conferences online creates a mix of effects, as forgoing in-person experiences benefits those who could not otherwise participate due to visa issues or travel costs\cite{Guinnessy:2021}. The immediate impact of cancelling a conference is readily apparent in pre-print counts in lattice physics (Figure~\ref{fig:1}; \arxiv\ subject code \texttt{hep-lat}, see Methods for explanation), where most pre-prints are posted around December each year as conference proceedings from the International Symposium on Lattice Field Theory. In 2020 the conference was cancelled\cite{LatticeConferenceWebsite} and no accompanying pre-prints exist. However, the impact on research from travel restrictions is likely to be much longer than what is represented by the drop in lattice physics pre-prints. Discussions at conferences or collaborative visits frequently spark new research ideas that might lead to a publication many months or years later.

Experimental research projects often require specialised laboratories, or data to be collected over many years. The pandemic has forced many laboratories to close or operate with restricted access\cite{PhysicsWorld,ScienceMag}, which could lead to long-lasting delays in ongoing experiments or immediate drops in publication rates in part due to difficulty in accessing completed (or nearly completed) experiment data. There is some evidence of this in pre-print numbers already, with {20\%} fewer pre-prints in high energy physics (\texttt{hep-ex}) in 2020 than expected by our model (observed 615; predicted 818 $\pm$ 94). But declines are not ubiquitous across experimental fields. The closest comparable field of phenomenology in high energy physics (\texttt{hep-ph}) showed no substantial decline. Neither did nuclear experimental research (\texttt{nucl-ex}). And throughout 2020, research in condensed matter (\texttt{cond-mat}) saw a $1.8\sigma$ increase above our model predictions (observed 16,188; predicted 15,325 $\pm$ 492). 
Segmenting condensed matter pre-prints by research topic shows that most of this increase (in \texttt{cond-mat}) was driven by a 30\% increase in material science pre-prints (\texttt{cond-mat.mtrl-sci}).

\begin{figure}
 \includegraphics[width=\linewidth]{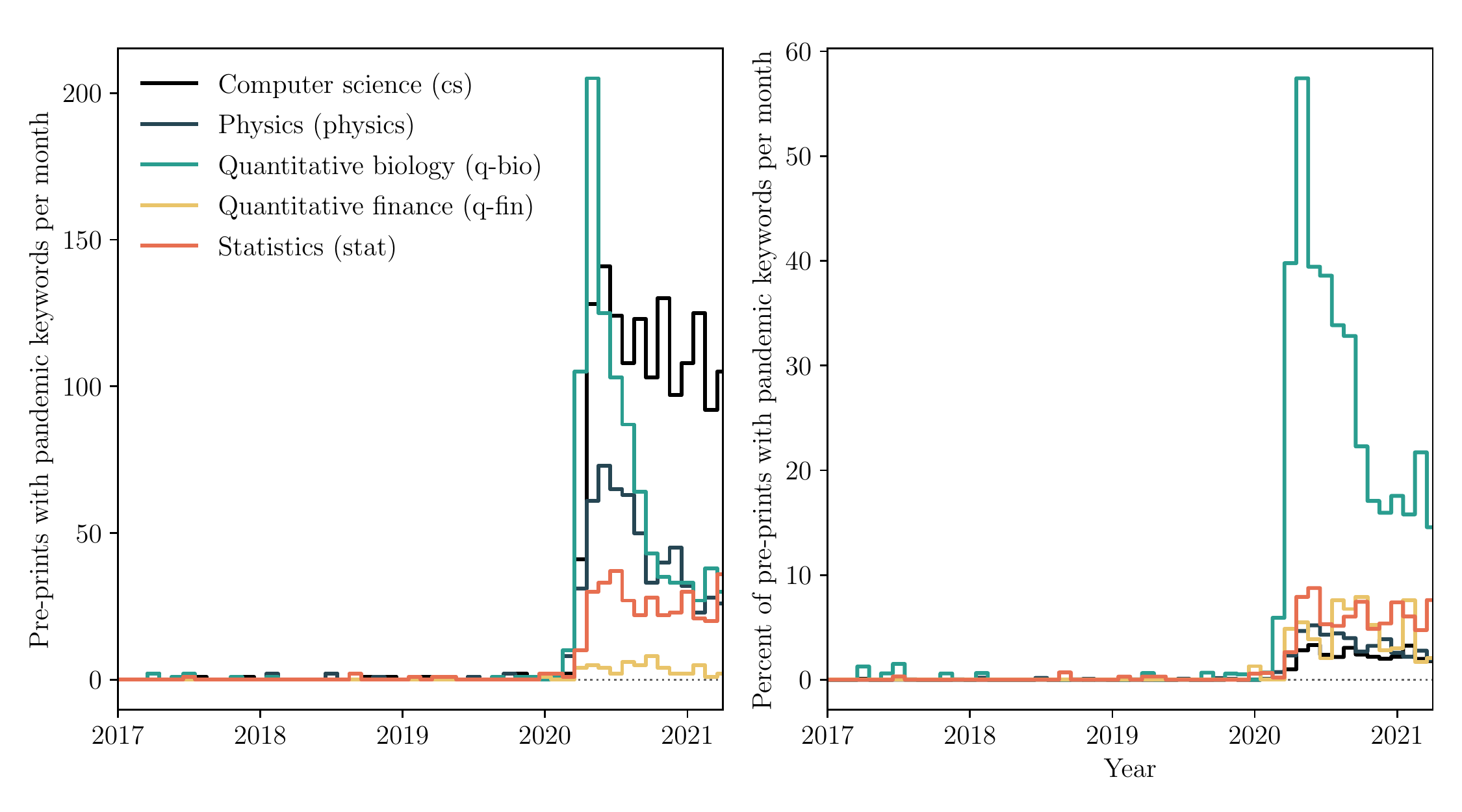} 
\caption{Pre-prints related to the COVID-19 pandemic peaked in April 2020, accounting for nearly 60\% the quantitative biology (\texttt{q-bio}) pre-prints in that month. Shown are five other research fields with the highest number of pandemic-related pre-prints in 2020. Monthly pre-prints are shown on the left, and the percent of pandemic-related pre-prints in that research field is shown on the right.}
\label{fig:2}
\end{figure}

The field of quantitative biology (\texttt{q-bio}) research showed the largest increase in pre-prints in 2020 above what is expected by our model. There were {2,790} quantitative biology pre-prints in 2020, {50\%} above the previous year, representing a $+8\sigma$ deviation from our model predictions (predicted 1,914 $\pm$ 109, 876 fewer than observed). This increase is explainable by an increase in COVID-19 related research, as there were 864 quantitative biology pre-prints in 2020 with pandemic-related terms in their title or abstract (see Methods), and just 39 in the decade prior. Indeed, in {April 2020} nearly 60\% of the \arxiv\ pre-prints in quantitative biology were related to the pandemic (Figure~\ref{fig:2}). Pandemic-related pre-prints also appeared in other fields (computer science, physics, statistics, and quantitative finance, all peaking around {April 2020}), but pandemic-related pre-prints constituted fewer than 10\% of the total pre-prints in these fields. As of mid-2021 the number of monthly pre-prints in quantitative biology has dropped to pre-pandemic levels. Approximately 20\% of current pre-prints in quantitative biology are related to the ongoing pandemic, representing a strong shift in ongoing research focus.

\renewcommand{\thefootnote}{$\dagger$} 
 
\begin{figure}[!h]
\begin{center}
	\includegraphics[width=0.95\linewidth]{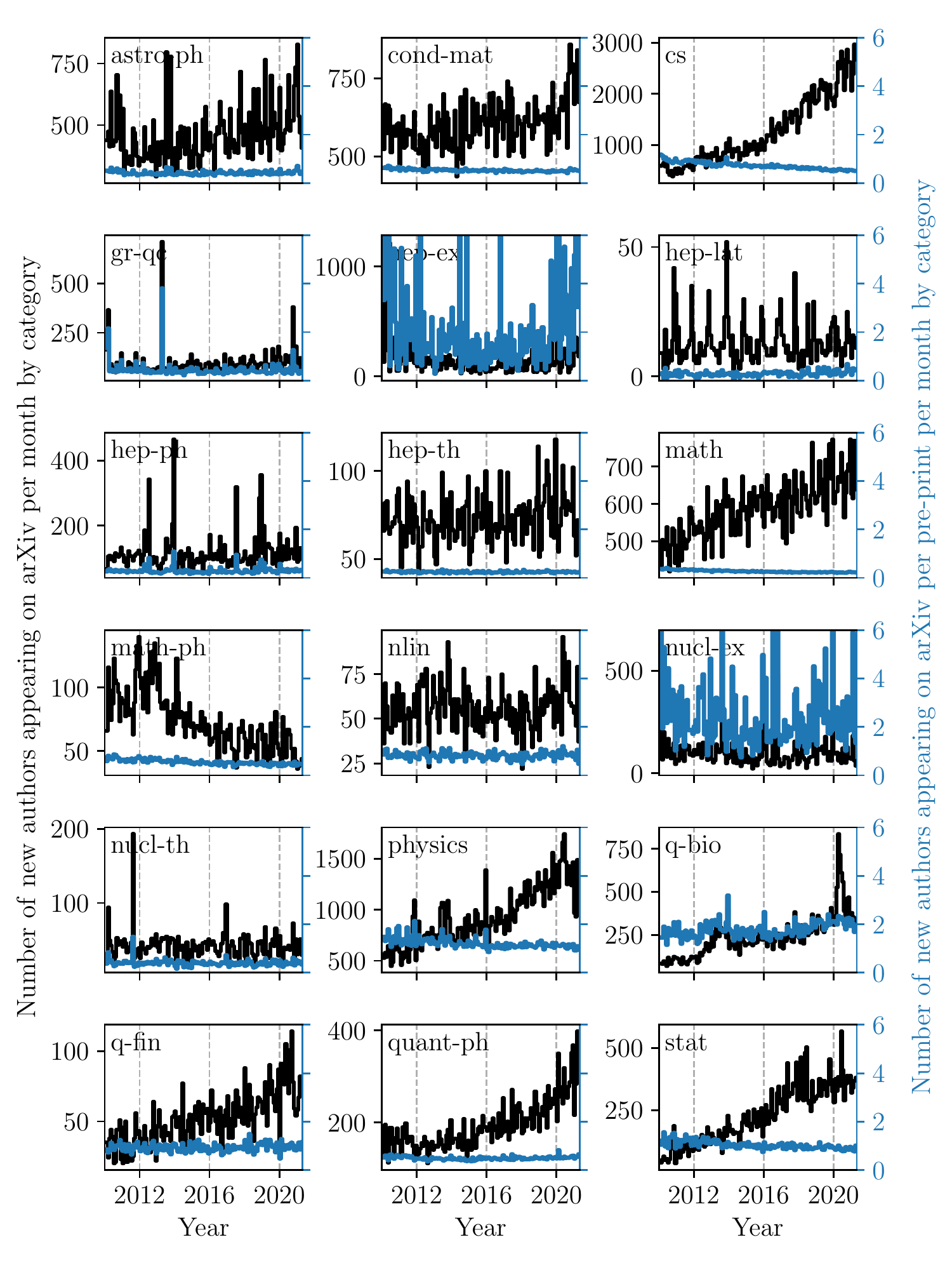} 
\end{center}
\caption{The number of new authors appearing in \arxiv\ pre-prints per month (black) peaked in quantitative biology in 2020. Here we show new authors since 2010, but we used data since 2007 to determine if an author was new. The slow change in new authors  approximates the net number of new authors joining the field. The number of new authors appearing per number of pre-prints is shown in light blue, with identical limits on the right-hand y-axis. This highlights authorship differences between fields (e.g., \texttt{hep-ex} and \texttt{hep-lat}) for large experiments.}
\label{fig:3}
\end{figure}

An increase in quantitative biology\footnote{Quantitative biology is only a sub-field of biology, and most biology pre-prints are posted to bioRxiv ({https://biorxiv.org}).} research during a global pandemic is unsurprising. However, the question arises if this increase was driven by existing researchers in the field, or by authors from other fields entering quantitative biology due to the pandemic. In 2020 there was a peak in the number of new authors appearing for the first time in the quantitative biology literature (Figure~\ref{fig:3}) while the number of new authors in all other fields remained steady. The increase in pre-prints (and new authors) is driven by small (1-4) groups of authors, where many had never posted pre-prints to quantitative biology before.

One quarter of the excess pre-prints posted to quantitative biology during 2020 were related to the pandemic and all authors had never appeared in the quantitative biology literature (234 of the 876 excess pre-prints in \texttt{q-bio}). A careful examination reveals that about half of these (45\%; 107/234) were written by mathematicians or physicists, with another 26\% (61/234) led by engineers, computer scientists, or economists. These pre-prints tend to focus on modelling the COVID-19 outbreak using public data sets, rather than quantitative biology research that requires more expert domain knowledge. About 3\% (8/234) were led by established biologists who have not used the \arxiv\ before and are now doing so presumably to help ensure that their COVID-19 research is more widely available. This fraction is likely higher among pre-prints where at least one author has appeared before in the quantitative biology literature. This represents a genuine increase in new researchers using the \arxiv, and a boost to making quantitative biology research more widely accessible. 







\begin{table}
       \caption{The number of \arxiv\ pre-prints per primary research field in 2020 compared to 2019, and model predictions (see Methods).}
\begin{center}
\small
    \label{tab:table1}
    \begin{tabular}{|c|l|c|c|c|c|} 
    \hline
    	\arxiv\ & Field of research & Pre-prints & Yearly & Predicted & Significance \\
	code &&(2020)& change (\%)&(2020)& ($\sigma$)\\
      \hline
\texttt{astro-ph}    & Astrophysics           & 14,893    & $ +3.4$ & 14,990 $\pm$ 653      & $-0.1$ \\
\texttt{cond-mat}    & Condensed Matter       & 16,188    & $ +7.0$ & 15,325 $\pm$ 492      & $+1.8$ \\
\texttt{cs      }    & Computer Science       & 54,808    & $+24.3$ & 51,673 $\pm$ 4,047    & $+0.8$ \\
\texttt{gr-qc   }    & General Relativity and & 3,405     & $+12.4$ & 3,166 $\pm$ 154       & $+1.6$ \\
					 & Quantum Cosmology      &&&&\\
\texttt{hep-ex  }    & High Energy Physics    & 615       & $-19.8$ & 818 $\pm$ 94          & $-2.2$ \\
                     & (Experiment)           &&&&\\
\texttt{hep-lat }    & High Energy Physics    & 394       & $-19.3$ & 475 $\pm$ 170         & $-0.5$ \\
                     & (Lattice)              &&&&\\
\texttt{hep-ph  }    & High Energy Physics    & 4,586     & $ +1.3$ & 4,615 $\pm$ 329       & $-0.1$ \\
                     & (Phenomenology)        &&&&\\
\texttt{hep-th  }    & High Energy Physics    & 3,869     & $ +0.9$ & 4,000 $\pm$ 291       & $-0.4$ \\
                     & (Theory)               &&&&\\
\texttt{math    }    & Mathematics            & 35,367    & $ +5.5$ & 35,362 $\pm$ 1,177    & $+0.0$ \\
\texttt{math-ph }    & Mathematical Physics   & 1,378     & $ -6.3$ & 1,460 $\pm$ 95        & $-0.9$ \\
\texttt{nlin    }    & Nonlinear Sciences     & 885       & $ +6.1$ & 822 $\pm$ 63          & $+1.0$ \\
\texttt{nucl-ex }    & Nuclear Experiment     & 499       & $ +2.9$ & 490 $\pm$ 61          & $+0.1$ \\
\texttt{nucl-th }    & Nuclear Theory         & 1,212     & $ -3.2$ & 1,234 $\pm$ 111       & $-0.2$ \\
\texttt{physics }    & Physics                & 15,241    & $+14.3$ & 14,746 $\pm$ 506      & $+1.0$ \\
\texttt{q-bio   }    & Quantitative Biology   & 2,790     & $+49.0$ & 1,914 $\pm$ 109       & $+8.0$ \\
\texttt{q-fin   }    & Quantitative Finance   & 979       & $+23.1$ & 805 $\pm$ 55          & $+3.2$ \\
\texttt{quant-ph}    & Quantum Physics        & 6,341     & $+13.4$ & 5,858 $\pm$ 266       & $+1.8$ \\
\texttt{stat    }    & Statistics             & 5,180     & $+20.2$ & 4,824 $\pm$ 490       & $+0.7$ \\
\hline
&\cellcolor{gray!25}\textbf{Total} & \cellcolor{gray!25}$\mathbf{168{,}630}$ & \cellcolor{gray!25}$\mathbf{+12.6}$ & \cellcolor{gray!25}$\mathbf{162{,}577 \pm 4{,}393}$ \cellcolor{gray!25}& \cellcolor{gray!25}$\mathbf{+1.4}$ \\
\hline 
    \end{tabular}
  \end{center}
\end{table}

The pandemic has also created distinct challenges when interpreting trends in \arxiv\ data. For example, there was a large spike in papers submitted to the \texttt{stat} category in June 2020 which one might assume was driven by COVID-related work given what we have observed in \texttt{q-bio}. However, the number of COVID-related papers posted in this period is modest, and the key factor appears to be delayed paper submission deadlines for two large conferences (NeurIPS and the Joint Statistical Meetings). Both conferences have had submission deadlines in mid-May in recent years, but in 2020, their deadlines were moved to early June.

The COVID-19 pandemic appears to have had minimal impact on collective research productivity to date. In this study we have only focussed on the quantity of publications as a measure for productivity, and not their quality. While citations are the most used measure of impact of academic publications, that metric becomes a more biased statistic when many related pre-prints are all being posted nearly at the same time\cite{Fassin:2021}. Amongst individual fields the largest decline is explainable by a cancelled conference, and the few other drops in pre-prints are statistically insignificant. It's plausible that many of the pre-prints already posted are research projects that were well underway before the pandemic. The relatively long timescales of academic research would suggest that the full impact on academic productivity due to the COVID-19 pandemic is yet to be seen. The approach applied here will be a useful quantitative approach to return to in subsequent years to explore the longer-term impact of the pandemic.

\section*{Methods}

\subsection*{Grouping by research field}

Pre-prints posted to the \arxiv\ can be listed in a single field of research, or cross-listed in multiple fields. A field is annotated by \texttt{primary.SEC} where the primary field of research has the prefix, and the sub-field of research is represented by a short suffix. For example, one pre-print may have a primary field of research as stellar astrophysics (\texttt{astro-ph.SA}) and be cross-listed in machine learning (\texttt{stat.ML}). These field(s) of research are supplied by the corresponding author. It is a subjective decision whether to include more than one field of research, or what those fields of research would be. For this reason, throughout this work when we segment by research field we take the primary parent research field provided and ignore any cross-listed fields of research. When we segment pre-prints by sub-field, we similarly take the primary sub-field provided and ignore any cross-listings.

A pre-print's identifier is defined by the year and month that the pre-print was posted, and the number of pre-prints already posted in that month (across all fields). Pre-prints posted prior to April 2007 use a different identifier scheme, and a different research field definition. For these reasons, we have excluded pre-prints earlier than this date.

\subsection*{Long-term modelling of monthly pre-print counts}

We use monthly pre-print counts per primary research field from January 2015 until December 2019 to make predictions for the number of pre-prints expected per month from January 2020 onwards. We use a straight line to model the mean pre-prints over time, and a Gaussian Process with a quasi-periodic kernel covariance function to capture seasonal periodicity\cite{Rasmussen:2006,Ambikasaran:2014}, and a white noise component. We fit the parameters of the mean model and the kernel hyper-parameters simultaneously, but kept the periodicity hyper-parameter fixed to one year. The predictions of the expectation and variance in the number of monthly pre-print counts are conditioned on the optimized hyper-parameters. We fit this model to monthly pre-print counts for every primary research field. These predictions are listed in Table~1. The expected total number of pre-prints, and the uncertainty in that number, are summed from the models used in individual research fields. 

Our conclusions are reasonably insensitive to the choice of model. As an alternative model we considered an autoregressive integrated moving average (ARIMA) model\cite{BoxJenkins:1970} with seasonal terms ($p=d=q=1$ for the non-seasonal component, and $P=D=Q=1$ with $m=12$ for the seasonal component). The predictions from the ARIMA model agree excellently with the mean predictions from the Gaussian Process model. The largest differences between these models is in \texttt{hep-lat}, where the ARIMA model predicts a higher peak number of pre-prints than the Gaussian Process model. This is not unexpected: the peak height of lattice physics pre-prints has been dropping for the last three years, and this gradual decline is better captured by the Gaussian Process model. If we adopted the ARIMA model then the drop in \texttt{hep-lat} pre-prints would become more significant. Throughout this work we adopt the predictions from the Gaussian Process model, unless otherwise specified.

\subsection*{Uniquely identifying authors}

The \arxiv\ team has parsed given and family names from author strings for all pre-prints. In general the author names have been appropriately parsed, even given a wide set of input formats. We made only one correction to these parsed names. When the author name was provided in the format of `Family-name First-given-name I.', without any commas, this was interpreted by the \arxiv\ parser such that the family name was `I.' and the given names were `Family-name First-given-name'. We identified situations like this by checking if the parsed family name was just an initial, and if two given names were parsed. In these situations, we correctly ordered the author name.

The \arxiv\ metadata available to us do not include institutional affiliations, or identifiers that would uniquely identify an author. 
There are two primary ways that name confusion could impact our inferences. In the first scenario, two people with the same name are amalgamated and treated as a single author that is on average twice as productive (or more, for very common names) as other authors. In the second scenario, an author will sometimes publish as `A.~B.~Smith', and other times publish as `A.~Smith'. A careful exploration of the data shows that this is a very frequent scenario, and if left uncorrected, would appear as many `unique' authors with half as many publications on average.

We have taken a simple approach to address name confusion. We first define a unique author by family name and the initial of the first given name (`Family-name, I.'), such that we intentionally group together authors that may share the same initial of their first given name. 
While our approach to name confusion is grossly simple, it is unlikely that these choices have any substantial impact on our inferences. In general, this errs on the side of name conflation and over-productivity, except for the relatively rare authors who change their last names over their career.
Any common name is likely to appear in the literature early in the data set and will not impact the conclusions we draw about how publishing changed in 2020. Indeed, we re-produced the analysis and figures in this paper by taking all initials given and it made no change to our conclusions. This is due, in part, because most authors do not provide middle initials.

\subsection*{Pre-prints related to the COVID-19 pandemic}

We identified a pre-print as being related to the COVID-19 pandemic if the title or abstract contained any of the following four (case-insensitive) terms: `pandemic', `COVID', `SARS-CoV-2', `lockdown', or `coronavirus'. Before 2020, these terms rarely appear in the title or abstract of any pre-print on \arxiv\ (Figure~\ref{fig:2}).

\vspace{\fill}

\ethics{No ethics approval was required for this work.}

\dataccess{The entire dataset used in this article is from \arxiv. The dataset is curated by Cornell University and hosted by Kaggle (https://www.kaggle.com/Cornell-University/arxiv). The dataset is updated weekly. We accessed this dataset on 01 June 2021.}

\aucontribute{All authors contributed to the discussions and have read and iterated upon the text of the final manuscript.}

\competing{The authors declare no competing interests.}

\funding{A.~R.~C. is supported in part by the Australian Research Council through a Discovery Early Career Researcher Award (DE190100656). 
Parts of this research were supported by the Australian Research Council Centre of Excellence for All Sky Astrophysics in 3 Dimensions (ASTRO 3D), through project number CE170100013.
I.~M. is a recipient of the Australian Research Council Future Fellowship FT190100574.}

\ack{We thank Peter Skands (Monash University) 
	 and Ross Young (University of Adelaide) 
for comments on publication trends in high energy physics.
}

\disclaimer{The paper reflects the authors' views and may not reflect those of their employers, funding agencies, or affiliated institutions.}

\pagebreak



\begin{thebibliography}{10}

\bibitem{Ginsparg:2011}
{Ginsparg}. P., {ArXiv at 20}, \emph{Nature}, \textbf{476}, 7359, doi:10.1038/476145a \newblock (2011).

\bibitem{Lariviere:2014}
{Larivi{\'e}re, V., et al.,} {arXiv E-prints and the journal of record: An analysis of roles and relationships}, \emph{Journal of the Association for Information Science and Technology}, \textbf{65}, 6, doi:10.1002/asi.23044 \newblock (2014).


\bibitem{Nicola:2020}
{Nicola, M., et al.,} {The socio-economic implications of the coronavirus pandemic ({COVID}-19): A review}, \emph{International Journal of Surgery}, \textbf{78}, 185--193, doi:10.1016/j.ijsu.2020.04.018 \newblock (2020).

\bibitem{Chu:2020}
{Yen-Hao Chu, I., Prima, A., Larson, H.~J., Leesa, L.}
{Social consequences of mass quarantine during epidemics: a systematic review with implications for the {COVID}-19 response}, \emph{Journal of Travel Medicine}, \textbf{27}, 7, doi:10.1093/jtm/taaa192 \newblock (2020).


\bibitem{Viglione:2020}
{Viglione, G.}, {Are women publishing less during the pandemic? Here?s what the data say}. \emph{Nature}, \textbf{581}, 356-366, doi:10.1038/d41586-020-01294-9 \newblock (2020).


\bibitem{Gewen:2020},
{Gewen, V.}, {The career cost of COVID-19 to female researchers, and how science should respond}. \emph{Nature}, \textbf{583}, 867--869, doi:10.1038/d41586-020-02183-x \newblock (2020).


\bibitem{IbnMohammed:2021}
{Ibn-Mohammed, T., et al.,} {A critical analysis of the impacts of {COVID}-19 on the global economy and ecosystems and opportunities for circular economy strategies}, \emph{Resources, Conservation and Recycling}, \textbf{164}, 105169, doi:1016/j.resconrec.2020.105169 \newblock (2021).

\bibitem{King:2021}
{King, M.~M., Frederickson M.~E.}, {The Pandemic Penalty: The Gendered Effects of COVID-19 on Scientific Productivity}. \emph{Socius}. January 2021. doi:10.1177/23780231211006977 \newblock (2021).  


\bibitem{Clement:2019}
{Clement, C.~B., Bierbaum, M., O'Keeffe, K.~P., Alemi, A.~A.,}
{On the Use of ArXiv as a Dataset},
Preprint at https://arxiv.org/abs/1905.00075 \newblock (2019).

\bibitem{Rasmussen:2006}
{Rasmussen, C.~E., Williams, C.~K.~I.,}
{Gaussian Processes for Machine Learning}, \emph{MIT Press}, ISBN 026218253X \newblock (2016).

\bibitem{Ambikasaran:2014}
{Ambikasaran, S., Foreman-Mackey, D., Greengard, L., Hogg, D.~W., O'Neil, M.,}
{Fast Direct Methods for Gaussian Processes}, Preprint at http://arxiv.org/abs/1403.6015 \newblock (2014).

\bibitem{Guinnessy:2021}
Guinnessy, P. (Physics Today). 2021. \emph{Physicists gathered virtually for APS March Meeting}. [online] Available at {https://physicstoday.scitation.org/do/10.1063/PT.6.3.20210405a/full} [Accessed 17 June 2021].
 
\bibitem{LatticeConferenceWebsite}
Helmholtz-Institut f\"ur Strahlen- und Kernphysik Indico Service (Indico). 2021. \emph{The 38th International Symposium on Lattice Field Theory (Lattice 2020)}. [online] Available at: {https://indico.hiskp.uni-bonn.de/event/1/} [Accessed 3 June 2021].

\bibitem{PhysicsWorld}
physicsworld (IOP Publishing). 2021. \emph{Critical research hit as COVID-19 forces physics labs to close}. [online] Available at {https://physicsworld.com/a/critical-research-hit-as-covid-19-forces-physics-labs-to-close/} [Accessed 4 June 2021].


\bibitem{ScienceMag}
Science Magazine (AAAS). 2021. \emph{Amid pandemic, Energy Department labs close to tens of thousands of users.} [online] Available at {https://www.sciencemag.org/news/2020/03/amid-pandemic-energy-department-labs-close-tens-thousands-users} [Accessed 4 June 2021].

\bibitem{Fassin:2021}
{Fassin, Y.} {Research on Covid-19: a disruptive phenomenon for bibliometrics}, \emph{Scientometrics}, \textbf{126}, 5305--5319, doi:10.1007/s11192-021-03989-w \newblock (2021).

\bibitem{BoxJenkins:1970}
{Box, G.~E.~P., Jenkins, G. M.} Time series analysis: Forecasting and control. San Francisco: Holden-Day. \newblock (1970).

\end{thebibliography}
\end{document}